\documentclass[aps,prl,twocolumn,groupedaddress, showpacs]{revtex4}

\usepackage{graphicx}
\usepackage{epsf,epsfig}
\usepackage{amssymb}

\newcommand{\be}{\begin{eqnarray}}
\newcommand{\ee}{\end{eqnarray}}
\newcommand{\ba}{\begin{array}}
\newcommand{\ea}{\end{array}}

\newcommand{\eps}{\varepsilon}

\newcommand{\erfc}{\mbox{erfc}}

\begin{document}

\title{Persistent current in small superconducting rings}

\author{Georg Schwiete$^1$ and Yuval Oreg$^{1,2}$}

\affiliation{$^1$ Department of Condensed Matter Physics, Weizmann
Institute of Science, Rehovot, 76100, Israel\\
$^2$ Department of Applied Physics, Stanford University, Stanford,
California, 94305, USA }

\date{\today}

\begin{abstract}
We study
theoretically the contribution of fluctuating Cooper pairs to the persistent
current in superconducting rings threaded by a magnetic flux. For
sufficiently small rings, in which the coherence length $\xi$
exceeds the radius $R$, mean field theory predicts a full reduction of
the transition temperature to zero near half-integer flux. We
find that nevertheless a very large current is expected to persist in
the ring as a consequence of Cooper pair fluctuations that do not condense. For larger
rings with $R\gg \xi$ we calculate analytically the susceptibility in the
critical region of strong fluctuations and show that it reflects
competition of two interacting complex order parameters.
\end{abstract}


\maketitle

\it Introduction and main results\rm - Superconducting fluctuations have been the
subject of intense research during the last decades~\cite{Larkin05}.
At temperature above the transition temperature $T_c$ to the superconducting state, when the system
is still metallic, pairs of electrons are formed for a limited time.
These superconduting fluctuations affect both transport and thermodynamic properties.

\begin{figure}[h]
\setlength{\unitlength}{2.3em}
\begin{picture}(12,5)
\put(0.5,0){\includegraphics[width=10.5\unitlength,height=5\unitlength]{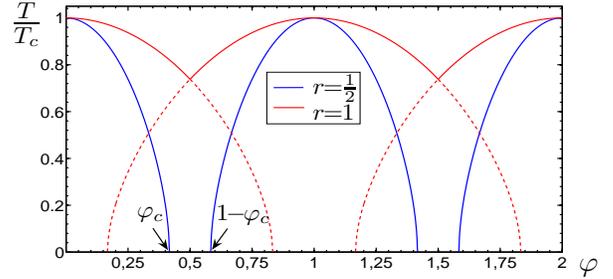}}
\end{picture}
\caption{Mean field (MF) phase diagram. $T_{c\varphi}$ separates the metallic (high $T$) and the superconducting (low $T$) phase as a function of the flux $\varphi=\phi/\phi_0$
through the ring. For sufficiently small rings with effective
radius $r=R/\xi < 0.6$ MF theory predicts a full reduction of
$T_c$ for fluxes between $\varphi_c \approx 0.83 r$ and
$1-\varphi_c$ near $\varphi=1/2$.
The transition line reflects the condition $\mathcal{L}^{-1}_{00}=0$, cf Eq.
(\ref{eq:fluctprop}).
 \label{fig:Tc}}
\end{figure}

 In bulk superconductors $T_c$ can be
reduced or even completely suppressed by various phase-breaking
mechanisms, for example by applying a magnetic field or
introducing magnetic impurities.
 A special situation occurs for superconducting rings and cylinders threaded by a magnetic flux
$\phi$. $T_c$ is \emph{periodically} reduced as a function of
$\phi$, a phenomenon known as Little-Parks
oscillations~\cite{Little62}. The period of the oscillations is
equal to~$1$ as a function of the reduced flux
$\varphi=\phi/\phi_0$, where the superconducting flux quantum
is $\phi_0=2\pi\hbar c/2e=\pi/e$ \cite{Units}, see
Fig.~\ref{fig:Tc}.

The magnitude of the maximal reduction in $T_c$ is
size-dependent. As we see in Fig.~\ref{fig:Tc}, mean field (MF)
theory predicts that for small rings or cylinders with $r\equiv
R/\xi<0.6$ the transition temperature is equal to zero in a
finite interval close to half-integer flux, giving rise to a
flux-tuned quantum phase transition, see  also
Eq.~(\ref{eq:fluctprop}) below. In this Letter we show that the
pair fluctuations give a large contribution to the persistent
current (PC) $I$ even at fluxes for which $T_c$ is reduced to
zero and the system has a finite resistance.

Recent experiments added significantly to our understanding of
fluctuation phenomena in superconductors with doubly-connected
geometry. Strong Little-Parks oscillations in the region where
$\xi >R$, where $T_c$ is reduced to zero,
 have been observed in a transport measurement on superconducting cylinders~\cite{Liu01}.
Koshnick et al.~\cite{Koshnick07} measured the PC in small
superconducting rings in the regime where  $R>\xi$, for the
smallest rings under study $T_c$ was reduced by $\approx
6\%$.
In this Letter we discuss the PC both in the regime of
moderate $T_c$ suppression for $r=R/\xi \gtrsim 1$  as well as
the strong Little-Parks oscillations for $r<0.6$. Before
presenting details of our approach, we summarize the main
results of our analysis.

\begin{figure}[h]
\setlength{\unitlength}{2.3em}
\begin{picture}(12,6.5)
\put(0.5,0.1){\includegraphics[width=10\unitlength,
height=6\unitlength]{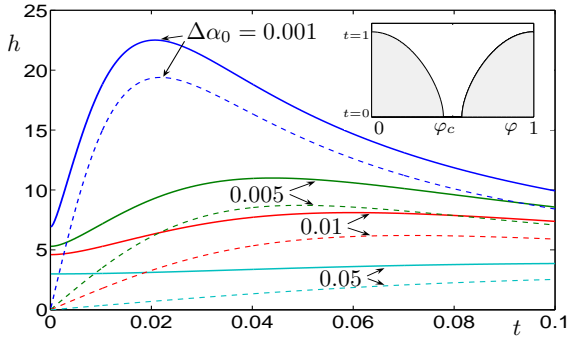}}
\end{picture}
\caption{The function $h$ that determines the PC close to the QCP,
$i_G\approx-\frac{2}{\gamma_E}\frac{1}{\varphi_c} h$, as a
function of $t=T/T_c$ for given $\Delta\alpha_0\approx
{2(\varphi-\varphi_c)}/{\varphi_c}$.
$\varphi_c=\frac{\pi}{2r\sqrt{2\gamma_E}}$ is the critical flux at
$T=0$ and $\gamma_E\approx 1.78$. $h$ is defined in the text. The dotted lines describe classical
fluctuations. Inset: MF phase diagram, superconducting region in
grey.} \label{fig:QCP}
\end{figure}

\paragraph{I. Regime with $r=R/\xi<0.6$:}
For $r<0.6$ the mean field $T_c$ vanishes and one would naively
expect a small normal state PC. We find, however, that close to
the critical mean field line (see Fig.~\ref{fig:Tc}) there is a
parametrically large enhancement of the PC due to quantum
fluctuations that decays only slowly away from that line. The
magnitude for the normal PC is $I_{N}\sim
\frac{1}{\phi_0}\frac{D}{R^2}\frac{1}{\log{g}}$, where $g$ is
the dimensionless ring conductance~\cite{Conductance08,Scurrent}. Our
calculations show that the PC due to pair fluctuations near the
critical flux $\varphi_c$ is parametrically larger and at low
$T$ given by
\be
I_{\rm FL}\approx - \frac{T_c}{\phi_0}\;\frac{1}{\varphi_c}\;\frac{\xi}{R}\log\left(\frac{1}{\Delta\varphi}\right)\label{eq:ifl},
\ee
where $\Delta\varphi\equiv (\varphi-\varphi_c)/\varphi_c$
measures the distance to the critical flux $\varphi_c$. When increasing $T$ the PC initially grows before going through a maximum at finite
$T$, where it can considerably exceed the result of Eq.~(\ref{eq:ifl}) [see Fig.~\ref{fig:QCP}]. Since $r^{-1}=\xi/R$ is a number of order 1 and $\frac{D}{R^2}=\frac{8}{\pi}\frac{T_c}{r^2}$ for a weakly disordered superconductor,
we find an enhancement factor of $\log(g)\log(1/\Delta\varphi)$.

Our results are obtained for the case when the flux acts as a
pair breaking mechanism. Other pair breaking mechanisms, e.g.
magnetic impurities or a magnetic field penetrating the ring
itself will lead to similar results. They cause a reduction of
$T_c$ to zero, the pair fluctuations, however, lead to a
parametric enhancement of the PC in the normal state.
Ref.~\cite{Bary08} suggests that a similar mechanism due to
magnetic impurities is related to the unexpectedly large PC in
noble metal rings \cite{Levy90, Ambegaokar90a}.

A metallic
state with small but finite resistance was observed experimentally
in superconducting cylinders~\cite{Liu01,Liuexperiment} with
$\varphi\approx 1/2$. Further theoretical and experimental studies
will be needed in order to clarify the relation to our findings,
where a large PC is caused by pair fluctuations that are unable to
condense.

\paragraph{II. Regime with $r>1$:}
The case $r>1$ is suitable for the description of the experiments
on persistent currents by Koshnick et al. \cite{Koshnick07}.
Previously the experiment has been interpreted using a
one-dimensional Ginzburg-Landau theory to describe the order
parameter fluctuations \cite{KoshnickConvergence}. Following these
lines one has to resort to numerical methods \cite{Oppen92} in
order to describe the critical region close to $T_c$, where
fluctuations proliferate.

 Our key observation is that part of the
rings in the experiment allow for a description using a suitable
generalization~\cite{Daumens98} of the $0d$ Ginzburg-Landau
theory. Indeed, following an expansion of the order parameter
field $\psi(\vartheta)$ in terms of angular momentum modes
$\psi_n$, a simple physical picture arises in the limit
$\sqrt{g}\gg r$. Two of the modes compete with each other close to
half-integer flux, while at the same time both of them strongly
fluctuate in the critical regime close to $T_c$.

Formally, the competition
arises due to the quartic term in the GL functional that induces
an interaction between the modes \cite{Imry74} and reveals itself in the
experiment mostly in the ``slope" of the PC, the susceptibility
$\chi=-\frac{\partial I}{\partial \phi}$. With this insight
$\chi$ can be calculated analytically even in the critical
fluctuation regime.

 As an example, denoting the susceptibility at
$T_c$ and zero flux by $\chi_0$ and at $T_{c,\varphi=1/2}$ by
$\chi_{1/2}$, we find
\be
\chi_{1/2}/\chi_0 \approx -2.7 \sqrt{g}/r. \label{eq:chi} \ee
Experimentally, a strong enhancement of the magnetic
susceptibility near $\varphi=1/2$ compared to $\varphi \approx
0$ was observed and Eq.~(\ref{eq:chi}) demonstrates that it is
controlled by the parameter $\sqrt{g}/r$. If it is large, the
current will rapidly change sign as a function of the flux at
half-integer flux, leading to a saw-tooth like shape of
$i_{\varphi}$. The full $T$ dependence of $\chi_{\varphi=1/2}$
is given in Eq.~(\ref{eq:f2}). For the smallest rings in Ref.~\onlinecite{Koshnick07,KoshnickUnits}, $\sqrt{g}\approx 33 r$.

\it Classical GL functional \rm - After presenting the main results in Eqs.~(\ref{eq:ifl}) and (\ref{eq:chi}) we now give more details of our
approach starting with the description of rings with only a moderate
suppression of $T_c$ (i.e. $r\gtrsim 1$).

When the superconducting coherence length $\xi(T)$ and the
magnetic penetration depth $\lambda(T)$ are much larger than the
ring thickness, the system is well described by a one-dimensional
order parameter field $\psi$ \cite{Imry02}. The partition function
can be written as a weighted average over configurations of the
order parameter $\psi$, $\mathcal{Z}=\int D\psi
\exp[-\mathcal{F}/T]$. Introducing angular momentum modes as
$\psi(\vartheta)=\frac{1}{\sqrt{V}}\sum_n\psi_n\;\mbox{e}^{in\vartheta}$,
where $V$ is the volume of the ring, the free energy functional
takes the form \be \mathcal{F}=\sum_n a_{n\varphi}|\psi_n|^2+
\frac{b}{2V}\sum_{nmkl}\delta_{n+k,l+m}\psi_n\psi^*_m\psi_k\psi_l^*.
\label{eq:fullF} \ee Here we wrote $a_{n\varphi}=\alpha
T_c\eps_{n\varphi}$, where $\eps_{n\varphi}=[T-T_{n\varphi}/{T_c}$
is the reduced temperature and
$T_{n\varphi}=T_c[1-(n-\varphi)^2/r^2]$ is determined by the sign
change of the coefficient $a_n(\varphi)$ and can thus loosely be
interpreted as the transition temperature of mode $\psi_n$
\cite{Microscopics08}. The mean field transition occurs at
$T_{c\varphi}$ that is equal to the maximal $T_{n}$ for given
$\varphi$, i.e. at the point where the first mode becomes
superconducting when lowering the temperature (cf. Fig.
\ref{fig:Tc}). The 0d Ginzburg parameter
$Gi=({2b}/{\alpha^2T_cV})^{1/2}$ is an estimate for the width of
the critical regime in the variable $\eps_n$. The parameter
$\sqrt{g}/r\approx 1/5r^2Gi$ has been used when stating our
results. Its relevance is now easily understood. $1/r^2$ is a
measure for the typical spacing between the transition
temperatures $T_n$ for different modes, since
$(T_0-T_1)/T_c=(1-2\varphi)/r^2$. This spacing should be compared
to the typical width of the non-Gaussian fluctuation region, $Gi$.
If it is large, a theory including only one or two
angular momentum modes is applicable.

\it Persistent current \rm - The persistent current $I$ is found
from the free energy $F=-T\ln \mathcal{Z}$ by differentiation
$I=-\partial F/\partial \phi$. The normalized current is given by
\be
\label{eq:Idef} i=I/(T_c/\phi_0)=\sum_{n=-\infty}^{\infty}
\frac{2\alpha}{ r^2}\left(n-\varphi\right)\left\langle
|\psi_n|^2\right\rangle.
\ee
 The averaging is performed with respect to the functional
$\mathcal{F}$ in Eq.~(\ref{eq:fullF}). $i_\varphi$ is periodic in
the flux $\varphi$ with period one. Since it is also an odd
function of the flux, it vanishes when the flux takes integer or
half-integer values.

\it Case $\varphi \approx 0$ \rm: The most important
contribution in the regime of non-Gaussian fluctuations close
to integer fluxes comes from the angular momentum mode $\psi_n$
with the highest transition temperature $T_{n\varphi}$. One may
then approximate Eq.~(\ref{eq:fullF}) by a single-mode and
calculate with $\mathcal{F}_n=
a_n|\psi_n|^2+\frac{b}{2V}|\psi_n|^4$ \cite{Buzdin02}. This is
the 0d limit of the GL functional~\cite{Muhlschlegel72} where
the functional integral becomes a conventional integral.
Indeed, performing the integral in polar coordinates, one finds
$\mathcal{Z}=\frac{\pi\sqrt{\pi}}{\alpha Gi}\exp(x_n^2) \
\mbox{erfc}(x_n)$, where $x_n=\eps_n/Gi$ \cite{Errorfunction}.
Using now Eq.~(\ref{eq:Idef}) with one mode only we find \be
i_n=4\Lambda(n-\varphi)f(x_n)\label{eq:scaling}\;\; \mbox{ for
} \varphi \approx n. \ee Here $\Lambda\equiv1/r^2Gi\approx
5\sqrt{g}/r$ and
$f(x)=\frac{\exp(-x^2)}{\sqrt{\pi}\mbox{{\scriptsize
erfc}}(x)}-x$ \cite{Errorfunction}. We note in passing the high
degree of universality implied by this result: All PC
measurements will fall on the same curve, if the PC -- measured
in suitable units $i =I/(T_c/\phi_0)$ -- and the reduced
temperature $\varepsilon_\varphi=(T-T_{c\varphi})/T_c$ are
scaled as $i \rightarrow i \frac{r}{\sqrt{g}}$,
$\epsilon_\varphi \rightarrow \epsilon_\varphi \ r \sqrt{g}$.
The scaling function $f$ was given above. This relation is a
valuable guide in characterizing different rings in
experiments.

Far above $T_c$ one obtains as a limiting case the Gaussian result
for a single mode $i_n\approx 2(n-\varphi)/r^2\eps_{n\varphi}$,
that can also be obtained directly by neglecting the quartic term
in the GL functional. It is known, however, that as soon as
temperatures are too high, $\eps_n\gg 1/r^2$, it is important to
sum the contribution of all modes~\cite{Ambegaokar90}. Far below
$T_c$ one recovers the mean field result $i_{MF} \equiv
\frac{-4}{r^2Gi^2}\eps_{n\varphi}(n-\varphi)$ for the PC in the
superconducting regime. An alternative route to finding the mean
field result would be to minimize
 the full single mode functional, which leads to the condition
$|\psi_n|^2=-a_nV/b$ and then to use Eq.~(\ref{eq:Idef}). The PC
$i_n$ in Eq.~(\ref{eq:scaling}) interpolates smoothly between the
Gaussian and the mean field result.

\begin{figure}[h]
\setlength{\unitlength}{2.3em}
\begin{picture}(12,6.5)

\put(0.5,0){\includegraphics[width=10.5\unitlength]{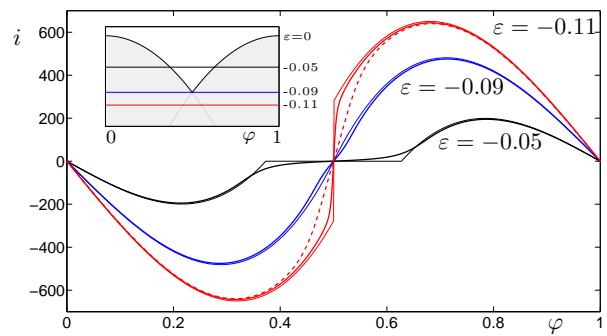}}

\end{picture}
\caption{ The PC $i=I/(T_c/\phi_0)$ as a function of the flux
$\varphi$. Parameters are $r=R/\xi=1.66$, $\Lambda=1/(r^2
Gi)\approx 5\sqrt{g}/r=50$, $\eps=(T-T_c)/T_c$. The transition for
$\varphi=1/2$ occurs at $\eps=-0.09$. Full lines: $i_2$ calculated
with $\mathcal{F}$ of Eq. (\ref{eq:2}), it takes into account two
modes and the interaction between them. We compare $i_2$ to two
approximations, which neglect this interaction. Dotted lines: The
mean field approximation $i_{MF}$ [discussed before
Eq.~(\ref{eq:2})] and dashed line: $i_{20}$ calculated with
$\mathcal{F}$ of Eq.~(\ref{eq:2}) \emph{without} coupling
\cite{DaumensMGauss}. Inset: MF phase diagram, superconducting
region in grey.} \label{fig:current}
\end{figure}

\it Case $\varphi \approx 1/2$: \rm A very interesting situation
occurs at half integer values of $\varphi$. The transition
temperatures for two modes become equal, their \emph{coupling}
becomes crucial ($\varphi\approx 1/2$ for definiteness), and we
approximate~\cite{Daumens98} \be \label{eq:2}
\mathcal{F}=\sum_{i=0,1}a_i|\psi_i|^2+\frac{b}{2\mbox{Vol}}\Big(|\psi_0|^4+|\psi_1|^4+4|\psi_0|^2|\psi_1|^2\Big).
\ee Calculation of the PC in the presence of the coupling requires
a generalization of the approach used for the single mode case
\cite{Daumens98,Persistentcurrent}. In Fig.~\ref{fig:current} we
display the PC $i_2$ as calculated from Eq. (\ref{eq:2}) for three
different temperatures, $T_{c\frac{1}{2}}<T<T_c$,
$T=T_{c\frac{1}{2}}$ and $T<T_{c\frac{1}{2}}$.  We compare it to
the MF result as well as to $i_{20}$ obtained by neglecting the
coupling $|\psi_0|^2|\psi_1|^2$ in Eq. (\ref{eq:2}).

 Above $T_{c\varphi}$ ($\varepsilon =-0.05$ in Fig.~\ref{fig:current}), in the region where the mean field result vanishes near half-integer flux,
the PC is \emph{purely fluctuational}.
We deduce from Fig.~(\ref{fig:current}) that the coupling of
the modes is crucial for $\chi(1/2)$, but not for the overall
shape when $T>T_{c\frac{1}{2}}$. However, just below
$T_{c\frac{1}{2}}$ ($\varepsilon =-0.11$ in
Fig.~\ref{fig:current}) the coupling is essential. The mean
field result is not applicable as it gives an infinitely sharp
jump in the PC at half-integer flux. The result without
coupling of the modes, $i_{20}$, gives a finite slope, but it
is far from the full current $i_2$ that includes the mode
coupling. The coupling drives the current $i_2$ towards the
mean field approximation $i_{MF}$ which includes only one mode.
This occurs because for a repulsive coupling the dominant mode
suppresses the subdominant one. Indeed, if mode $n=0$ is
dominant then the coupling adds a mass term $2b/V
\left\langle\left|\psi_0^2 \right|\right\rangle
\left|\psi_1\right|^2$ to mode $n=1$ and reduces its $T_c$.

\it Susceptibility \rm - We will now discuss in more detail the
slope at half-integer flux, which is most sensitive to the
coupling between the modes below, and to the non-Gaussian
fluctuations close to $T_c$.
Differentiating the expression \cite{Persistentcurrent} for $i_2$
we obtain \be
\overline{\chi}_{\varphi=1/2}&=&\chi/(T_c/\phi_0^2)=4\Lambda\;g_1\left(x\right)-4\Lambda^2\;g_2\left(x\right),
\label{eq:f2} \ee $x=\frac{\eps}{Gi}+\frac{1}{4}\Lambda$ \cite{Xpar08}. The
dimensionless smooth functions $g_1(x)=\frac{1}{2J(x)}\textrm{e}^{\frac{1}{3}x^2}\textrm{erfc}(x)-\frac{2x}{3}$ and
$g_2(x)=\frac{3}{2\sqrt{\pi}J(x)}\;\textrm{e}^{-\frac{2}{3}x^2}-\frac{3x}{2J(x)}
\textrm{e}^{\frac{1}{3}x^2}\;\textrm{erfc}(x)-1$, where $J(x)=\int_x^\infty dt\;\textrm{e}^{\frac{1}{3}t^2}\;\textrm{erfc}(t)$, obey
$g_1(0)\approx 0.78$ and
$g_2(0)\approx0.315$.
For large $\Lambda =1/r^2Gi\approx 5\sqrt{g}/r$ one can neglect
the first term in Eq.~(\ref{eq:f2}). Then one obtains
$\overline{\chi}_{1/2}=-\;4\Lambda^2\; g_2\left(x\right)$. For
the susceptibility close to integer flux one easily obtains
$\overline{\chi}_{0}=4\Lambda
f(x_0)$  from Eq.~(\ref{eq:scaling}). Comparing to the expression for
$\overline{\chi}_{1/2}$, we find Eq.~(\ref{eq:chi}).

This is the
strong enhancement of $\chi_{1/2}$ compared to $\chi_0$ observed
in the experiment~\cite{Koshnick07}.
Comparison with the numerical calculation of Ref.
\cite{Koshnick07} shows that our analytical results are accurate to within a
few percent already for $\sqrt{g}/r\gtrsim 8$
\cite{KoshnickUnits,KoshnickConvergence}.

\it Quantum critical regime \rm - So far we have discussed the
limit $r=R/\xi >1$, where the suppression of $T_c$ is small and
a finite temperature phase transition occurs. We will now
discuss the case where $r=R/\xi <0.6$ and $T_c$ is reduced to
zero at a critical flux $\varphi_c$ near $\varphi=1/2$, see
Fig.\ref{fig:Tc}. Near the quantum critical point (QCP) it is
no longer legitimate to use the classical GL functional, in
which only the static component of the order parameter field is
considered. Instead, all Matsubara frequencies should be taken
into account in the imaginary time formalism. The full
fluctuation propagator is given by: \be
(\nu\mathcal{L})_{nk}^{-1}=\ln\left[\frac{T}{T_c}\right]+\psi\left[\frac{1}{2}+\frac{\alpha_n+|\Omega_k|/2}{2\pi
T}\right]-\psi\left[\frac{1}{2}\right],\quad\label{eq:fluctprop}
\ee where
$\alpha_n(\varphi)=\frac{1}{2}\frac{D}{R^2}(n-\varphi)^2$ and
$\Omega_k=2\pi kT$ are bosonic Matsubara
frequencies~\cite{Galitski01,Lopatin05}. Following the standard
approach, we first find the critical line $\varphi(T)$ in the
temperature-flux plane by equating $\mathcal{L}_{00}^{-1}=0$.
For the QCP at $T=0$ one obtains the critical flux
$\varphi_c=\pi r/(2\sqrt{2\gamma_E})$, $\gamma_E\approx 1.78$
\cite{Units}. Due to the flux-periodicity of the phase
diagram, the QCP can only be observed in the ring geometry if
$\varphi_c<1/2$ which implies $r<\sqrt{2\gamma_E}/\pi\approx
0.6$. Notice that this critical value of $r=R/\xi$ is 20\%
larger
than a naive application of the quadratic approximation valid for $r
\gg 1$ would suggest.

Restricting ourselves to the interval $\varphi\in(0,0.5)$  we
find that near the QCP it is sufficient to consider the $n=0$
mode. In the Gaussian regime we obtain (cf. Fig~\ref{fig:Tc}) the
following fluctuation contribution to the PC
$i_G=-\frac{2\nu\varphi}{T_c\varphi^2(T)}\;T\sum_k\mathcal{L}_{0k}$ \cite{general}.
Expanding $\mathcal{L}_0^{-1}$ in small
$\Delta\alpha=[\alpha_0(\varphi)-\alpha(\varphi(T))]/\alpha_0(\varphi_c)$
we find with logarithmic accuracy
$i_G=-\frac{16\varphi}{\pi^2r^2}h(\Delta\alpha,t)
\stackrel{\mbox{\tiny${\varphi
\!\!\rightarrow\!\!\!\varphi_{\!c}}$}}{\longrightarrow}
 -\frac{2}{\gamma_E}\frac{1}{\varphi_{c}}
h(\Delta\alpha,t)$,
$h(\Delta\alpha,t)=\ln\frac{s}{\Delta\alpha}+\frac{1}{2s}-\psi(1+s)$,
$s=\frac{\Delta\alpha}{2\gamma_E t}$ and $t=T/T_c$.

 A few remarks are in order concerning this result. The second
term in the expression for $h$ is the classical $\Omega=0$
contribution to the sum. The upper cut-off for the frequency
summation has been chosen as
$\overline{\Omega}=2\alpha_0(\varphi(T))$ \cite{cutoff}. The function $h$ has
the asymptotic form $ h
\approx{\gamma_Et}/{\Delta\alpha}+\ln({1}/{2\gamma_E t})$ for
$\Delta\alpha\ll t\ll 1$ and $h\approx\ln{1}/{\Delta\alpha}$
for $t\ll \Delta\alpha\ll1$. It is important that
$\Delta\alpha$ is $T$-dependent and in order to reveal the full
$T$-dependence of $i_G$ one should first find the transition
line $\alpha_0(\varphi(T))$. $h(T)$ is displayed in
Fig.~\ref{fig:QCP}. The maximum of $h$ at finite $T$ is a
result of two competing mechanisms. As $T$ grows from zero,
thermal fluctuations become stronger. At the same time the
distance to the critical line becomes larger for fixed
$\varphi$, which eventually leads to a decrease of $i_G$.

\it Conclusion \rm - In conclusion, we showed that on the
normal side of the flux-tuned superconductor normal-metal
transition in small rings the fluctuation PC can be very large
compared to the normal case and decays only logarithmically
away from the critical point. For larger rings as studied in recent experiments we obtained detailed analytical
predictions for the strong fluctuation region.

\begin{acknowledgments}
We thank E.~Altman, H. Bary-Soroker, A. I. Buzdin, A. M. Finkel'stein, Y.~Imry,
Y.~Liu, F. von Oppen for useful discussions, and K. Moler, N.
Koshnick and H. Bluhm for stimulating discussions and for sharing
their numerical results with us. We acknowledge financial
support from the Minerva Foundation, DIP and ISF grants.
\end{acknowledgments}

\begin{thebibliography}{10}
\bibitem{Liuexperiment}
In Ref.~\cite{Liu01} the resistance $R$ along the cylinder drops as $T$
  decreases and saturates for the lowest $T$. It has been suggested that the
  observable regime is dominated by thermal fluctuations \cite{Lopatin05} and
  that at even lower $T$ an upturn of $R$ could be expected. For the PC these
  regimes correspond to $T$ above and below the maximal PC. The role of
  inhomogeneities along the cylinder axis, that has no analog for rings, was
  stressed in Ref. \cite{Vafek05}.
\bibitem{KoshnickConvergence}
For rings with $\sqrt{g}\gg r$, below $T_c(\varphi\approx 1/2)$ the numerical
  procedure converges very slowly and no comparison to the
  experimental data was performed in Ref.~\cite{Koshnick07}.
\bibitem{KoshnickUnits}
  The rings with largest $\Lambda$ displayed in \cite{Koshnick07} have
  $\Lambda\approx 70$, but rings with larger $\Lambda$ have been measured.
\bibitem{Persistentcurrent}
The result is $i_2=-4\frac{r^2}{Gi}\mathcal{M}_-+\frac{8}{3}x_-\mathcal{M}_+$,
  where
  $\mathcal{M}_{\pm}=x_{\pm}\mp\frac{1}{\mathcal{P}}\left(\textrm{e}^{x_0^2}\;%
\textrm{erfc}(x_0)\pm \mbox{e}^{x_1^2}\;\erfc(x_1)\right)$ and
  $\mathcal{P}=4\int_0^\infty dz\;
  \textrm{e}^{3z^2+2(2x_1-x_0)z+x_1^2}\;\textrm{erfc}(2z+x_1)$,
  $x_{\pm}=(x_0\pm x_1)/2$. An analogous formula has been given in Ref.
  \cite{Daumens98}.
\bibitem{general} In general: $i_G=-T/T_c\sum_{nk}\mathcal{L}_{nk}\partial\mathcal{L}^{-1}_{nk}/\partial\varphi$. Further away from the QCP $i_G$ should be evaluated numerically.
\bibitem{cutoff}
The divergence is cured by the contribution of all non-singular modes with $n\ne 0$.
\end{thebibliography}

\end{document}